\documentclass[journal=jpclcd,manuscript=letter,layout=twocolumn]{achemso} 
\setkeys{acs}{usetitle=true, doi=true}

\usepackage{chemformula} 
\usepackage[T1]{fontenc} 

\usepackage{amsmath,amssymb}
\usepackage[version=3]{mhchem} 
\usepackage{graphicx}
\graphicspath{{./figures/}}

\usepackage{bm} 

\usepackage{hyperref}
\hypersetup{colorlinks=true, linkcolor=blue,urlcolor=blue!50!black, citecolor=blue!50!black}

\usepackage{cleveref}
\crefrangelabelformat{equation}{(#3#1#4)$-$(#5#2#6)}

\usepackage[separate-uncertainty]{siunitx}

\usepackage{chemarr}
\usepackage{cuted}
\usepackage{setspace}

\usepackage[normalem]{ulem}

\usepackage{lmodern}

\author{\normalsize Arthur V. Straube}
\affiliation{
\normalsize Zuse Institute Berlin, Takustra{\ss}e 7, 14195 Berlin, Germany}
\email{straube@zib.de}
\author{Stefanie Winkelmann}
\affiliation{
\normalsize Zuse Institute Berlin, Takustra{\ss}e 7, 14195 Berlin, Germany}

\author{Christof Sch\"utte}
\affiliation{
\normalsize Zuse Institute Berlin, Takustra{\ss}e 7, 14195 Berlin, Germany}
\alsoaffiliation{\normalsize Freie Universit{\"a}t Berlin, Department of Mathematics and Computer Science, \\ Arnimallee 6, 14195 Berlin, Germany}

\author{Felix H{\"o}fling}
\affiliation{\normalsize Freie Universit{\"a}t Berlin, Department of Mathematics and Computer Science, \\ Arnimallee 6, 14195 Berlin, Germany}
\alsoaffiliation{\normalsize Zuse Institute Berlin, Takustra{\ss}e 7, 14195 Berlin, Germany}

\title[]{\Large Stochastic pH oscillations in a model of 
	the urea–urease reaction confined 
	to lipid vesicles\footnote{
  This document is the unedited author’s version of a submitted work that was
  subsequently accepted for publication in The Journal of Physical Chemistry Letters, copyright \copyright American Chemical Society after peer review. To access the final edited and published work,
  see DOI: \href{https://doi.org/10.1021/acs.jpclett.1c03016}{10.1021/acs.jpclett.1c03016}  \\[3mm] \textsf{\textbf{Cite as:}  A.~V.~Straube, S. Winkelmann, C. Sch\"utte, \\F. H\"ofling, \textit{J. Phys. Chem. Lett.} {\bf 12}, 9888--9893 (2021)} } }

\begin{document}

\begin{tocentry}
\includegraphics[width=1.0\textwidth]{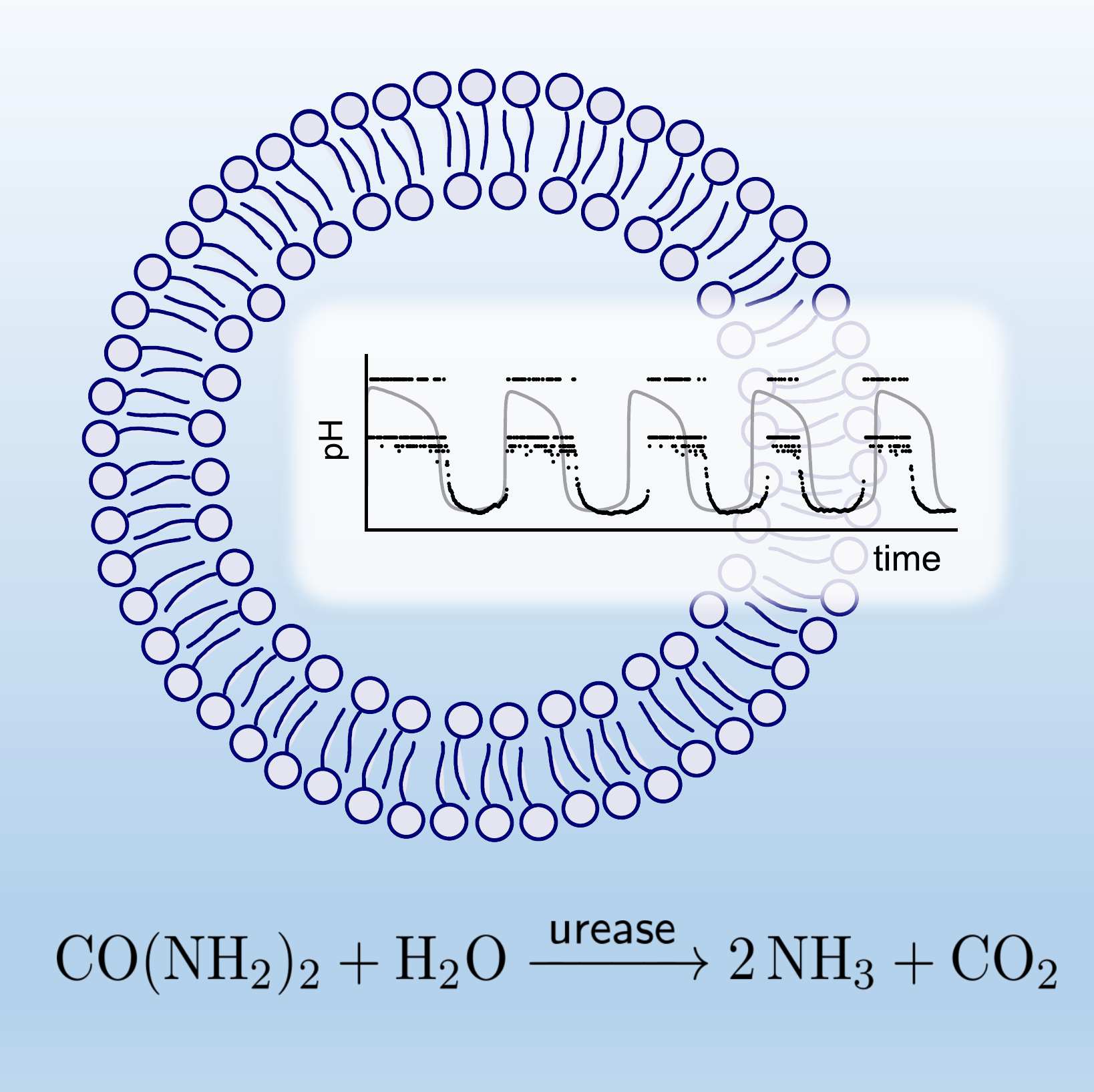}
\end{tocentry}

\begin{abstract}
The urea--urease clock reaction is a pH switch from acid to basic that can turn into a pH oscillator if it occurs inside a suitable open reactor.
We study the confinement of the reaction to lipid vesicles, which permit the exchange with an external reservoir by differential transport, enabling the recovery of the pH level and yielding a constant supply of urea molecules.
For microscopically small vesicles, the discreteness of the number of molecules requires a stochastic treatment of the reaction dynamics.
Our analysis shows that intrinsic noise induces a significant statistical variation of the oscillation period, which increases as the vesicles become smaller.
The mean period, however, is found to be remarkably robust for vesicle sizes down to approximately \SI{200}{\nano\meter}.
The observed oscillations are explained as a canard-like limit cycle that differs from the wide class of conventional feedback oscillators.
\end{abstract}


Oscillations are vital for the processes of life such as metabolism, signalling, cell growth and division \cite{Goldbeter:book1996, Novak:NRMCB2008} with examples ranging from fast signalling cycles and calcium oscillations to slow circadian rhythms\cite{Panda:Nature2002}.
Cells gain control over these processes by biochemical reaction networks \cite{Tyson:COCB2002, Alon:book2020}, e.g., gene-regulatory, protein-interaction, and metabolic networks,
which almost always involve enzyme-catalyzed reactions.
Protonation and bi-protonation can significantly affect the enzymatic activity, leading to a bell-shaped dependence of the reaction speed on the \ch{H+} concentration or, equivalently, the pH level \cite{Alberty-Massey:BBA1954}.
Such a dependence can give rise to pronounced periodic pH variations, the key driving factor for pH oscillators \cite{Orban:ACR2015}.
A conventional \ch{pH} oscillator
is built up by balancing a positive, autocatalytic feedback (production of \ch{H+}) with a time-delayed, negative feedback (e.g., consumption of products) \cite{Novak:NRMCB2008, Orban:ACR2015}. 
A qualitatively different pH oscillator has recently been suggested for a lipid vesicle with the urea-urease clock reaction\cite{Lente-etal:NJC2007, Hu-etal:JPCB2010, Bubanja-etal:RKMC2018} periodically recovered by the differential transport  of acid and urea across the vesicle membrane \cite{Bansagi:JPCB2014, Miele:Proc2016, Miele:LNBE2018}.

Experimentally, urea--urease pH oscillations were observed so far in macroscopic reaction volumes \cite{Hu-etal:JPCB2010, Muzika-etal:PCCP2019}. 
Also, most analyses of pH oscillators to date have relied on deterministic reaction rate equations (RRE). 
Furthermore, there is a growing interest in chemical oscillators for applications.\cite{Cupic:FC2021,Shklyaev:FC2020,Maria:FC2020, Budroni-etal:CC2021, Mallphanov-Vanag:PCCP2021}
This motivates the question whether stable limit cycles persist and how they change upon
downscaling from the macroscopic to, e.g., intracellular reaction volumes.
Indeed, the cytoplasm is a highly heterogeneous medium exhibiting macromolecular crowding and compartmentalization, with repercussions on the reaction kinetics \cite{Hoefling:RPP2013, Weiss:2014, Schneider:B2015,Tsiapalis:B2021}.
Enzymatic activity is confined to small reaction chambers
ranging from about \SI{10}{\micro\meter} for lipid membrane organelles \cite{Cooper-Hausman:2009book} down to
\SI{20}{\nano\meter} for bacterial microcompartments \cite{Sutter-etal:NSMB2008, Kerfeld-etal:ARM2010} and outer membrane vesicles \cite{KaparakisLiaskos-Ferrero:NRI2015, Schwechheimer-Kuehn:NRM2015}.
Such small compartments can host only very limited copy numbers of molecules, necessitating the replacement of RREs by their discrete and inherently stochastic counterparts \cite{Grima-Schnell:EB2008, Wilkinson:NRG2009, Winkelmann:2020book}.
Intrinsic noise due to such molecular discreteness leads to a breakdown of the macroscopic theory of Michaelis--Menten kinetics
\cite{Stefanini:NL2005, Grima:PRL2009, Grima:BMCSB2009}.
For monostable reaction networks, not only the size of fluctuations\cite{Grima-etal:JCP2011, Thomas-etal:BMCSB2012} but also the mean concentrations \cite{Thomas-etal:JCP2010, Ramaswamy-etal:NC2012} become volume dependent. Furthermore, intrinsic noise may change the stability of steady states, inducing oscillations in deterministic systems without limit cycles \cite{McKane-etal:JSP2007, Thomas-etal:JTB2013}, or alter the characteristics of limit cycles \cite{Ramaswamy-Sbalzarini:SR2011}.
However, its impact on pH-regulated systems has remained largely unexplored.

In this work, we consider the urea--urease reaction and study how the stable rhythmic variation of the pH level \cite{Miele:Proc2016, Muzika-etal:PCCP2019}
is affected by intrinsic noise when decreasing reaction volumes to biologically relevant scales.
Within one cycle, molecular copy numbers can vary from few molecules to several thousands almost instantaneously, which is captured by the stochastic simulations. 
We detect irregular oscillations, perform a statistical analysis of the period lengths and gain further insight into the oscillation mechanism.

\begin{figure*}[tb]
	\centering
	\includegraphics[width=1.00\textwidth]{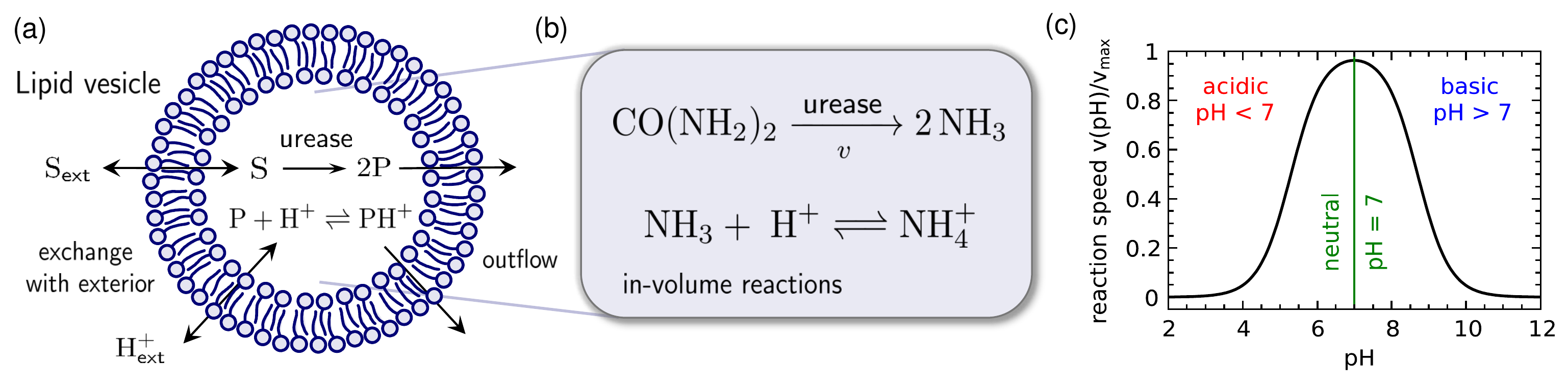}
	\caption{\textbf{Schematic representation of the 
	enzyme-assisted reaction network.} (a) The enzyme (urease) catalyzes conversion of the substrate (urea) into product (ammonia) in a lipid vesicle compartment affected by changing acidity (hydrogen ion, \ch{H+}). The products (ammonia and ammonium) are subjected to decay or outflow from the vesicle, while the substrate and acid exchange with the exterior of the vesicle. (b) Volume reactions \eqref{4sm-R1} and \eqref{4sm-R2} taking place in the vesicle, showing the meaning of the involved components; \ch{H_2O} and \ch{CO_2} have been omitted in the first reaction.~
	(c) The reaction speed $v=k_\mathrm{cat}([\ch{S}],[\ch{H}^+])\cdot [\ch{S}]$ of the catalytic step \eqref{4sm-R1} evaluated for urease, see \cref{eq:kcat}, 
	shows a strong dependence on the level of $\ch{pH}=-\log_{10}([\ch{H+}]/\SI{1}{M})$; it is fastest in a neutral medium ($\ch{pH} \approx 7$).
	}
	\label{fig:sketch}
\end{figure*}

\textit{Model.} Our study is based on a minimal model for the urease-catalyzed urea hydrolysis, which exhibits pH oscillations
while admitting a simple representation as a reaction network to facilitate the stochastic simulations.
B\'ans\'agi and Taylor\cite{Bansagi:JPCB2014} showed that the full model of the urea--urease reaction cycle, involving the concentrations of eight molecular species, can be reduced to an effective 5-variable model.
To further simplify, we eliminate one more species (\ch{OH-}) from the reaction network with merely small quantitative changes to the evolution of the remaining concentrations (see Supporting Information).
The corresponding reaction scheme involves only four species and consists of two core reactions
that are assumed to take place inside a lipid vesicle, serving as a small-size, well-mixed reaction compartment of volume $V$.
In addition, the vesicle can exchange molecules with its exterior via a permeable membrane (Figure~\ref{fig:sketch}a).
Under the action of urease enzymes, urea \ch{CO(NH_2)_2} as the substrate species \ch{S} is converted into ammonia molecules \ch{NH_3} as product \ch{P} (Figure~\ref{fig:sketch}b). Concomitantly, ammonia reacts with the acid to form ammonium ions (abbreviated as \ce{PH+} in the following).
Thus, the reactions inside the vesicle read:
\begin{subequations}\label{4sm}
	\begin{gather}
	\ch{S}  \xrightarrow[{k_\textrm{cat}}]{\text{\scriptsize\,\sf urease\;}} \ch{2 P}\,, \label{4sm-R1} \\
	\ch{P + H+} \xrightleftharpoons[k_{2r}]{k_{2}} \ch{PH+}\,. \label{4sm-R2}
	\end{gather}
\end{subequations}
The speed of reaction \eqref{4sm-R1} is crucially affected by the acidity of the medium and controlled by the available amount $X_{\ce{H+}}$ of protons \ch{H+}; the proton concentration $[\ch{H+}]=X_{\ch{H+}}/V_\textrm{M}$ is converted to the pH value via
$\ch{pH}=-\log_{10}([\ch{H+}] / \SI{1}{M})$
in terms of the molar volume $V_\textrm{M}=V N_\textrm{A}$ and Avogadro's number $N_\textrm{A}$.
Hereafter, we denote the numbers of molecules of species \ch{S}, \ch{H+}, \ch{P}, and \ch{PH+} as $X_{\ch{S}}$, $X_{\ch{H+}}$, $X_{\ch{P}}$, $X_{\ch{PH+}}$, respectively; we will reserve square brackets to refer to concentrations, $[\ch{S}] = X_{\ch{S}} / V_\textrm{M}$, etc.

The efficacy of the catalytic step \eqref{4sm-R1} is modeled by an effective rate\cite{Alberty-Massey:BBA1954,Fidaleo-Lavecchia:CBEQ2003, Bansagi:JPCB2014}
\begin{align}
k_\textrm{cat}([\ch{S}],[\ch{H+}])
 & = \frac{k_\textrm{cat}^\textrm{M}([\ch{S}]) } {
	 1+  [\ch{H+}]/K_\textrm{E1} +K_\textrm{E2} / [\ch{H+}] }   \,, \label{eq:kcat}
\end{align}
with the conventional Michaelis--Menten rate in the absence of \ch{pH}-effects given by
\begin{align}
k_\textrm{cat}^\textrm{M}([\ch{S}]) & = \frac{v_\textrm{max} } {
	K_\textrm{M}+[\ch{S}]
} \label{eq:kcat-MM}
\end{align}
and the Michaelis--Menten constant\cite{Krajewska:JMCBE2009,Hu-etal:JPCB2010, Bansagi:JPCB2014} $K_\textrm{M}=\SI{3e-3}{M}$. 
The rate $k_\textrm{cat}([\ch{S}],[\ch{H+}])$ possesses a maximum that is proportional to $v_\textrm{max}$ at an optimal amount of \ch{H+}, and reaction \eqref{4sm-R1} is suppressed for \ch{H+} concentrations smaller and larger than this value, or, equivalently at large and small \ch{pH} values, as determined by the constants \cite{Krajewska:JMCBE2009,Hu-etal:JPCB2010, Bansagi:JPCB2014} $K_\textrm{E1}=\SI{5e-6}{M}$ and $K_\textrm{E2}=\SI{2e-9}{M}$, see Figure~\ref{fig:sketch}c.
For the stochastic simulations in terms of particle numbers $X_{\ch{S}}$ we evaluate $k_\textrm{cat}(X_{\ch{S}}/V_\textrm{M},X_{\ch{H+}}/V_\textrm{M})$ as the propensity for reaction \eqref{4sm-R1} to occur. 
This reaction is further coupled to reaction \eqref{4sm-R2}, meaning that the product is also affected by the acidity and can reversibly turn into ammonium ions \ch{PH+}. The corresponding rates are set as\cite{Eigen:ACIE1964,Hu-etal:JPCB2010, Bansagi:JPCB2014} $k_2=\SI{4.3e10}{M {\tothe{-1}} {\s}\tothe{-1}}$ and $k_{2r}=\SI{24}{\s\tothe{-1}}$.

Apart from the in-volume reactions, \eqref{4sm-R1} and \eqref{4sm-R2}, we assume outflow or decay of the product in both its forms, \ch{P} and \ch{PH+}, with the rate constant $k>0$. 
Further, we consider an explicit exchange of \ch{S} and \ch{H+} with the exterior of the vesicle serving as a reservoir, with rates $k_\textrm{S}$ and $k_\textrm{H}$, respectively, equal in both directions. The spatial exchange between the interior and the exterior of the vesicle is modeled as stochastic jump process along the lines of the spatio-temporal master equation \cite{winkelmann2016spatiotemporal,winkelmann2021mathematical} and can be written as reactive transitions. Thus, the interaction with the exterior of the reaction volume is summarized as
\begin{subequations}\label{24sm-out}
	\begin{gather}
	\ch{P } \xrightarrow[]{\, k \,} \varnothing\,, \quad 
	\ch{PH+ } \xrightarrow[]{\, k \,} \varnothing\,, \label{4sm-outp} \\
	\ch{S} \xrightleftharpoons[]{\,k_\mathrm{S}\,} \ch{S}_\textrm{ext}\,, 
	\quad 
	\ch{H+} \xrightleftharpoons[]{\,k_\mathrm{H}\,} \ch{H+}_\textrm{ext}\,. \label{24sm-exch} 
	\end{gather}
\end{subequations}
We treat the reservoir as sufficiently large, such that reactions \eqref{24sm-exch} lead only to marginal changes to the amounts of $\ch{S}_\textrm{ext}$ and $\ch{H+}_\textrm{ext}$. Therefore, we approximate their concentrations by fixed values $[\ch{S}_\textrm{ext}]$ and $[\ch{H+}_\textrm{ext}]$ and replace reactions \eqref{24sm-exch} by
\begin{align}\label{24sm-exch2}
\ch{S} \xrightleftharpoons[\,{k_\mathrm{S} [\ch{S}_\textrm{ext}]}]{k_\mathrm{S}} \varnothing\,,
\quad
\ch{H+} \xrightleftharpoons[\,{k_\mathrm{H} [\ch{H+}_\textrm{ext}]}]{k_\mathrm{H}} \varnothing\,.
\end{align}

\begin{figure*}
	\centering
	\includegraphics[width=0.6\textwidth]{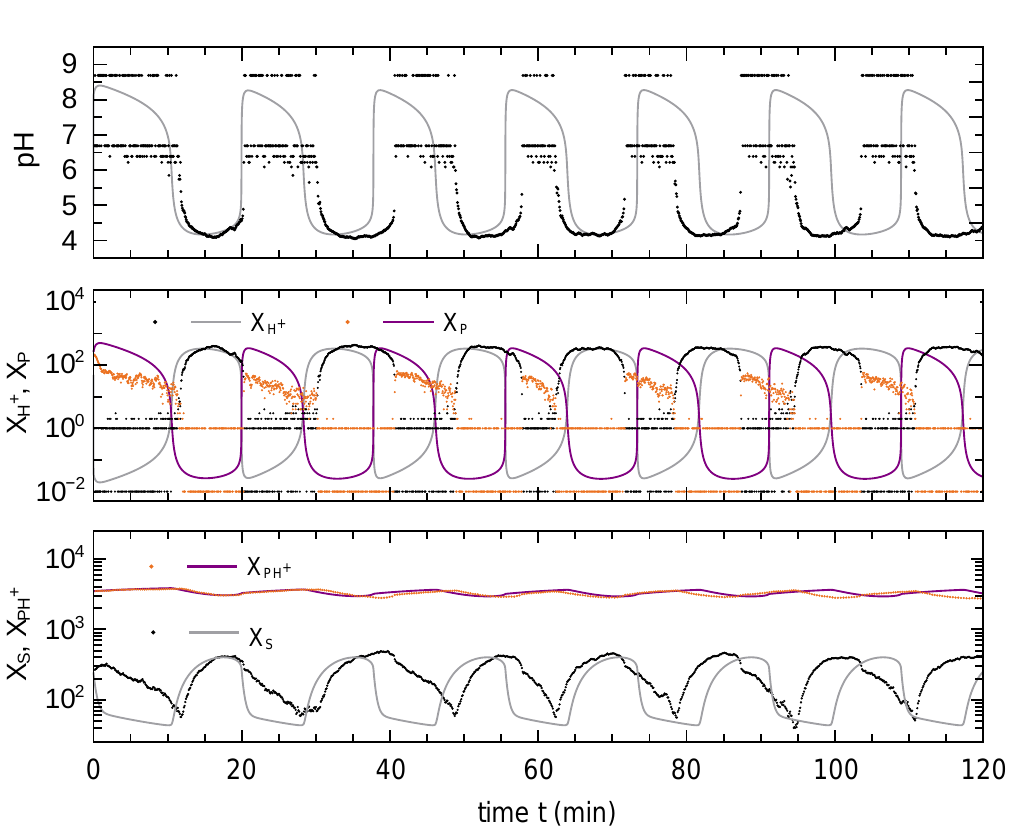}
	\caption{\textbf{Stochastic evolution of molecule numbers.}
	The \ch{pH} level (upper panel) and molecule numbers for the species \ch{H+} and \ch{P} (middle panel) and \ch{S} and \ch{PH+} (bottom panel) as functions of time for the urea--urease reaction scheme [Figure~\ref{fig:sketch} and \cref{4sm,eq:kcat,eq:kcat-MM,24sm-out,24sm-exch2}] confined to a vesicle of \SI{250}{\nano\meter} in diameter, which corresponds to a reaction volume of $V=\SI{8.18e-18}{\liter}$. Note the logarithmic scales.
	Solid lines are solutions to the deterministic reaction rate equations [Supporting Information, eqs.~\eqref{eq:rre}].
	Dots show an exemplary solution to the stochastic reaction dynamics obtained with Gillespie's stochastic simulation algorithm \cite{Gillespie:JPC1977}; the special value $X_{\ce{H+}} = 0$ is represented as $0.01$ ($\ch{pH}\approx 8.7$).
	}
	\label{fig:evol-4sm-250nm}
\end{figure*}

To inspect oscillatory regimes, we rely on the parameters that were shown to exhibit periodic deterministic oscillations for the urease-loaded membrane \cite{Bansagi:JPCB2014}. Generally, the rate of proton transport $k_\textrm{H}$ shall be faster than that of urea $k_\textrm{S}$; here, we use $k_\textrm{H} =\SI{9e-3}{\s\tothe{-1}}$ and $k_\textrm{S} = \SI{1.4e-3}{\s\tothe{-1}}$. The outflow rate of the products is set to $k = k_\textrm{S}$ and the maximum speed as $v_\textrm{max} =\SI{1.85e-4}{M\, \s^{-1}}$, where the latter corresponds to an urease concentration of \SI{50}{U}.
In all simulations, the external values of $X_{\ch{S}}$ and $X_{\ch{H+}}$ were fixed to match the concentrations $[\ch{S}_\textrm{ext}]=\SI{3.8e-4}{M}$ and $[\ch{H+}_\textrm{ext}]=\SI{1.3e-4}{M}$, or equivalently, to an acidic environment at $\ce{pH}=3.9$.
Inside of the vesicle, the initial values of $X_{\ch{S}}$ and $X_{\ch{H+}}$ were chosen to correspond to concentrations $[\ch{S}]_0 =\SI{5e-5}{M}$ and $[\ch{H+}]_0=\SI{e-5}{M}$ (or $\ce{pH}=5$), respectively.

\textit{Results and discussion.}
The deterministic evolution of the macroscopic concentrations obeys the RREs of the 4-species model (see eqs.~\eqref{eq:rre} of the Supporting Information).
For the parameter values chosen above, the results from numerical integration are quantitatively similar to the earlier findings within the 5-variable model \cite{Bansagi:JPCB2014}.
For an exemplary vesicle size of \SI{250}{\nano\meter} in diameter (i.e., a reaction volume of $V=\SI{8.18e-18}{\liter}$) the evolution of all variables after a short transient  displays clear periodic oscillations (Figure~\ref{fig:evol-4sm-250nm}, solid lines).
Especially, the pH level varies strongly between $\ch{pH} \approx 4.2$ and $\ch{pH} \approx 8.3$ (upper panel).
Correspondingly, the copy number of protons $X_{\ch{H+}} =[\ch{H+}] V_\textrm{M}$ (as rescaled solution of the RRE) on a logarithmic scale mirrors this behavior,
and the product $X_{\ch{P}}$ evolves in antiphase relative to $X_{\ch{H+}}$, with both quantities changing over four orders of magnitude rapidly (middle panel).
The values of $X_{\ch{PH+}}$ show comparably little variation and remain large and distinctly greater than those of the other species (bottom panel).
The maximum copy numbers of $X_{\ch{P}}$ and $X_{\ch{H+}}$ are similar in magnitude to the typical values for the substrate \ch{S}, while the minima of $X_{\ch{P}}$ and $X_{\ch{H+}}$ correspond formally to average copy numbers of the order \num{e-2}. Although such values are not prohibited by the reaction rate formalism, the actual copy numbers must be integer with the closest allowed values being either $0$ or $1$. This inconsistency is a  signature of the deficiency of the macroscopic description at such a small scale.

\begin{figure*}
	\centering
	\includegraphics[width=1.0\textwidth]{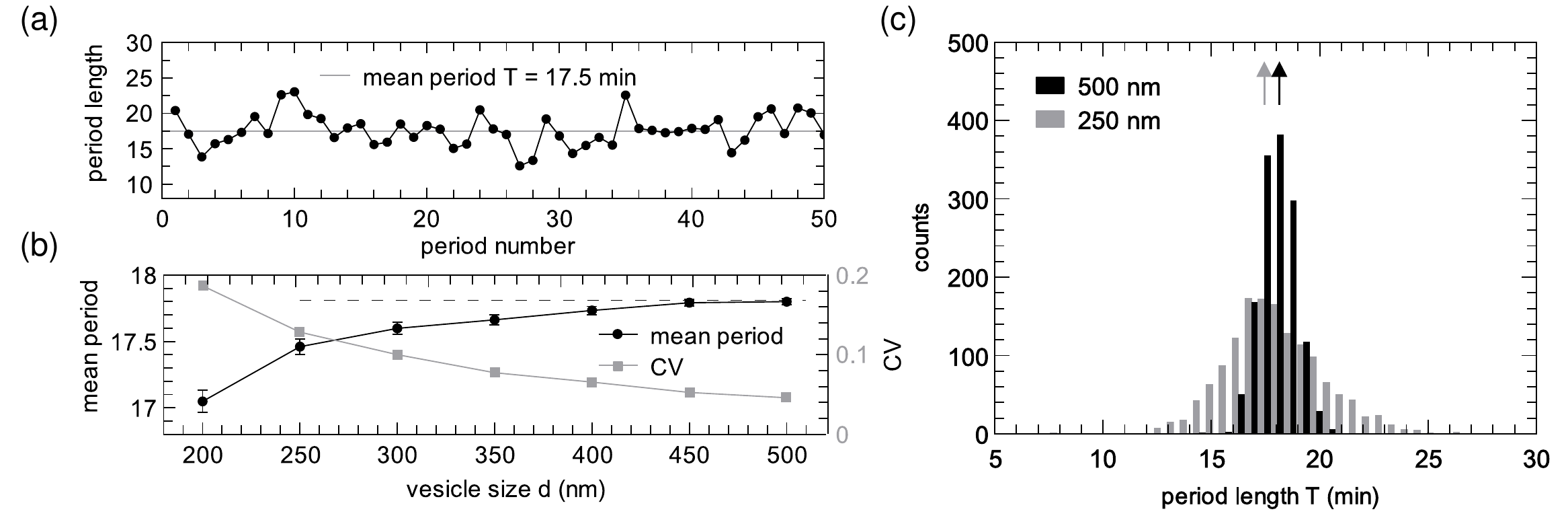}
	\caption{\textbf{Time periods and their statistical characteristics.} (a) Sequence of period lengths (dots) for a vesicle size of \SI{250}{\nano\meter} (reaction volume $V =\SI{8.18e-18}{\liter}$) with the mean over $1500$ periods (horizontal line).~
	(b) Mean periods with their statistical errors (disks and bars) and coefficient of variation (CV, squares) for different vesicle sizes.
	The dashed line indicates the macroscopic value of the period length, $T_\textrm{det}=\SI{17.8}{\minute}$.~
	(c) Histogram of the period lengths for vesicle sizes of \SI{500}{\nano\meter} and \SI{250}{\nano\meter};
	arrows indicate the mean periods. 
	}
	\label{fig:TimePeriods}
\end{figure*}

Stochastic simulations of the reactions \eqref{4sm}, \eqref{4sm-outp}, and \eqref{24sm-exch2} were performed by the stochastic simulation algorithm \cite{Gillespie:JPC1977, Gillespie-etal:JCP2013}.
In the macroscopic limit of a large reaction volume (e.g., for giant vesicles of diameter \SI{10}{\micro\meter}), the stochastic concentrations converge to the solution of the deterministic RRE given by eqs.~\eqref{eq:rre}, as expected.\cite{Gillespie-etal:JCP2013, Winkelmann:2020book}
With decreasing volume, the role of intrinsic noise grows and one anticipates deviations from the deterministic description.
The stochastic trajectories develop well pronounced fluctuations and differ significantly from the corresponding deterministic solutions,
as demonstrated for a vesicle size of \SI{250}{\nano\meter} in Figure~\ref{fig:evol-4sm-250nm}.
These stochastic effects are weaker for species of large copy number, e.g. \ch{PH+}, while they are strong for the acid and the product, whose amounts drop to few molecules and even becomes zero frequently.
The same features are reflected in the oscillations of the pH level, which directly follows from $X_{\ch{H+}}$.

We stress that the intrinsic noise perturbs the rhythm of the pH variation.
The stochastic oscillations become clearly irregular, from time to time showing either longer or shorter periods compared to their strictly regular deterministic counterparts.
To characterize this kind of stochasticity, we have extracted the period lengths $T$ from a single, long trajectory of $X_{\ch{H+}}$ covering about 1500 periods. The obtained sequence of $T$-values shows a high variability (Figure~\ref{fig:TimePeriods}a) around the mean period of $T_\text{av} = \SI{17.46+-0.06}{\minute}$, which is slightly shorter than the value predicted by the deterministic model, $T_\text{det} = \SI{17.79+-0.01}{\minute}$.
Further, the data show no sign of a temporal trend in the period length, and an autocorrelation analysis suggests that the lengths of subsequent periods are independent.
The large scatter of period lengths along a stochastic trajectory is evidenced from their statistical distribution, shown in  Figure~\ref{fig:TimePeriods}c for vesicle sizes of \SI{250}{\nano\meter} and \SI{500}{\nano\meter}. The scatter is larger for the smaller vesicle, and we infer a small shift of the mean value.
Indeed, Figure~\ref{fig:TimePeriods}b corroborates that the mean oscillation period becomes monotonically shorter upon decreasing the size of the vesicle.
At the same time, the coefficient of variation (CV), which is the dimensionless ratio of the standard deviation over the mean,
gradually grows for smaller reaction volumes, see Figure~\ref{fig:TimePeriods}(b).
At large volumes, CV tends to zero as required by the macroscopic limit;
for the smallest vesicle size shown (\SI{200}{\nano\meter}), we have $\text{CV} \approx 0.2$.
Generally, this trend is expected since smaller reaction volumes correspond to more discrete and therefore more noisy systems. Overall, with the decrease in volume, the oscillations become more and more irregular.
For very small vesicles (e.g., \SI{100}{\nano\meter}, see Fig.~S1 of Supporting Information), the size of fluctuations becomes comparable to the oscillation amplitude and the oscillatory behavior breaks down.

\begin{figure*}
	\centering
	\includegraphics[width=\textwidth]{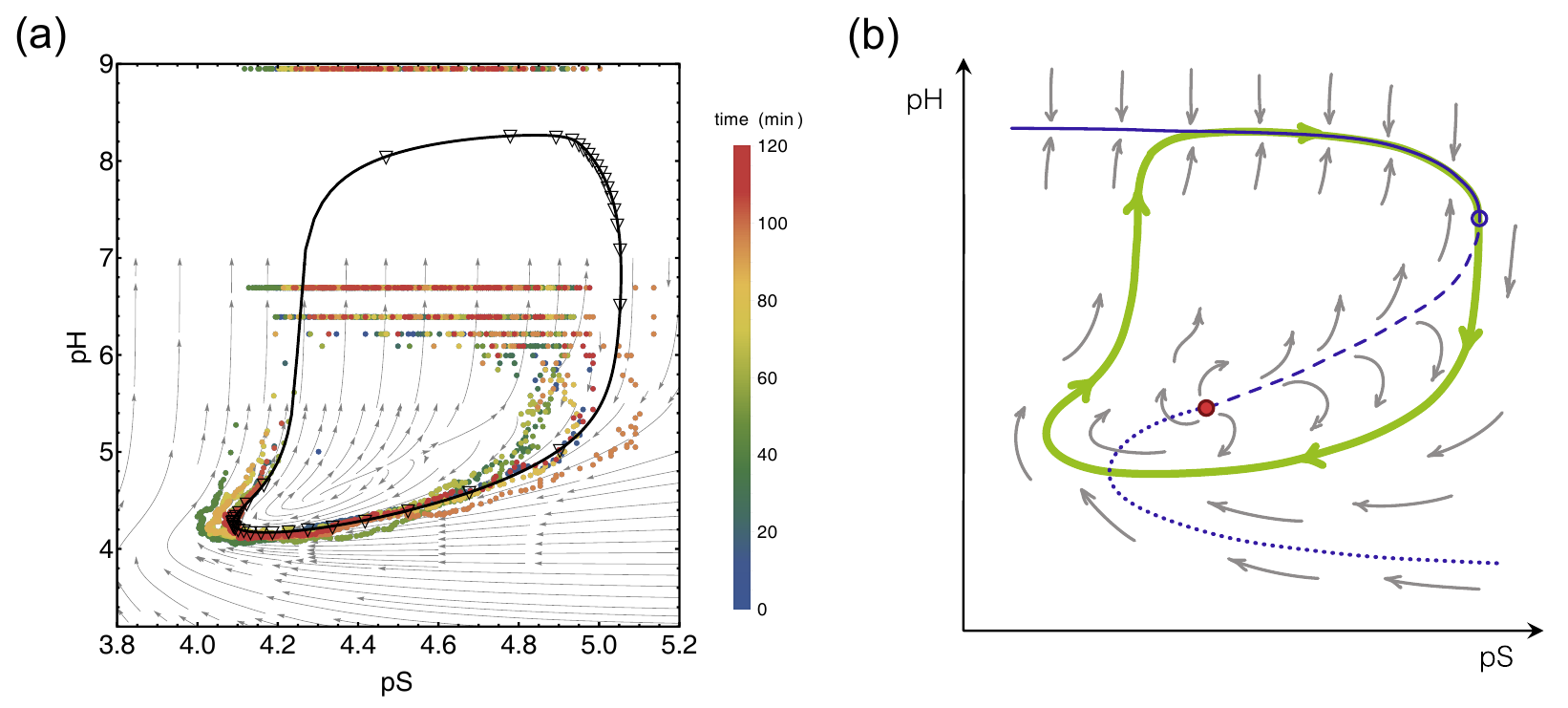}
	\caption{\textbf{Phase portrait and limit cycle in the \ch{pH}-\ch{pS} plane.} (a) Limit cycle from the deterministic model (solid line) and stochastic simulations (dots) 
		with $\ch{pS}=-\log_{10}(X_{\ch{S}}/V_\textrm{M})$.
		Stochastic results are for a vesicle size of \SI{250}{\nano\meter} and cover 6 oscillation periods (same data as in Figure~\ref{fig:evol-4sm-250nm}); data points with $X_{\ce{H+}}=0$ are drawn at the upper frame border, and colors encode time.
		Open triangles along the limit cycle are equally spaced in time and indicate the speed along the cycle (clockwise);
		their distribution reflects the alternating phases of fast and slow motion.
		Grey arrows depict the flow of the dynamic system, obtained approximately for a reduced, 2-variable model (see main text).
		~(b) The structure of the flow and the limit cycle (thick green line) emerge from the combination of an unstable focus (solid, red circle) and canard-type behavior. The latter is determined by the \ch{pH}-nullcline (purple line), where $d[\ch{H+}]/dt=0$, which consists of an attractive branch (solid line), passing through a turning point (open circle) to a repelling (dashed line) and a neutrally stable (dotted line) branch.
		Grey arrows indicate the direction of the phase flow.
	}
	\label{fig:PhasePlot}
\end{figure*}

For a dynamic system showing regular oscillations, the deterministic solution (after an initial transient) follows a limit cycle, i.e., an attractive, closed orbit in the space of concentrations.
For the 4-variable model studied here, Figure~\ref{fig:PhasePlot}a shows the deterministic limit cycle (solid line) in the pH-pS plane, where $\ch{pS} = -\log_{10} (X_{\ch{S}}/V_\textrm{M})$, overlaid with a short exemplary stochastic trajectory (dots).
In this representation, the cycle is followed clockwise.
We infer that intrinsic noise causes pronounced irregularities of the stochastic loop, with the trajectory points distributed well around the deterministic cycle for the smaller values of \ch{pH} (high $X_{\ch{H+}}$), but significantly deviating from it for larger \ch{pH} (low $X_{\ch{H+}}$).
The latter is due to the fact that non-integer copy numbers are not permitted in the stochastic simulation.
The discreteness of the number of protons $X_{\ch{H+}}$ is apparent in the figure for $\ch{pH} \gtrapprox 6$
and incompatible with the deterministic solution, which implies $0 < X_{\ch{H+}} < 1$ for $\ch{pH} \gtrapprox 7$ for the chosen reaction volume.
Thus, the lowest possible values of $X_{\ch{H+}}$ either undershoot ($X_{\ch{H+}} = 1$) or overshoot ($X_{\ch{H+}} = 0$, i.e., formally $\ch{pH}=\infty$) the upper branch of the deterministic limit cycle.

Further insight into the oscillation mechanism is gained by studying the structure of the deterministic flow (Figure~\ref{fig:PhasePlot}a, stream lines).
As such a flow map is non-trivial to obtain for a system of more than two variables, we have approximately reduced the $4$-variable RRE system to a two-dimensional dynamic system, by resorting to the quasi-steady-state assumptions\cite{Segel-Slemrod:SIAM-Rev1989} for the products $\ch{P}$ and $\ch{PH+}$. This \emph{ad-hoc} simplification preserves the fixed points of the original system and captures the qualitative structure of the flow; in particular, it yields a limit cycle quantitatively close to that of the full model for $\ch{pH} \lesssim 6.5$. The closed-loop attractor results from the interplay of an unstable focus point at $(\ch{pS}, \ch{pH}) \approx (4.31, 4.57)$, which is the only fixed point, and a canard-type behavior\cite{Benoit-etal:CM1981, Desroches-Jeffrey:PRSA2011} at high \ch{pH} values. This combination leads to an oscillator motif that differs from standard pictures of bistability \cite{Orban:ACR2015, Novak:NRMCB2008, Hirsch:DynamicalSystems} (Figure~\ref{fig:PhasePlot}b).

Typical for canard-type behavior is a coupling between fast and slow dynamics.
As depicted in Figure~\ref{fig:PhasePlot}b, the upper branch of the \ch{pH}-nullcline (i.e., the manifold $d[\ch{H+}]/dt = 0$) is strongly attractive (solid line) and combines fast, almost transverse motion towards the limit cycle orbit followed by the creeping along it (\ch{pS} increases), which holds until the turning point (open circle) is reached.
At this point, the nullcline bends back accompanied by a change of stability: the manifold between the turning point and the fixed point (red circle) is unstable (dashed line) and locally separates the flow into regions of increasing and decreasing \ch{pH} level.
In contrast to toy models for canard dynamics, where the dynamics switches between two attractive branches, the rest of the nullcline in the present system is neutrally stable (dotted line) and has no obvious effect on the flow structure.
Thus, after reaching the turning point, the pH level decreases quickly and the phase trajectory follows the flow set between the repelling manifold and the outer flow field, around the unstable focus until the loop is closed.
For the stochastic trajectories, which traverse the interior of the limit cycle, we infer that different crossing points of the separatrix lead to a scatter in the phase plane (near $\ch{pS}\approx 4.8{-}5.1$), which explains the observed variability in the period length.

In conclusion, we have studied the urea--urease reaction confined to a nanosized lipid vesicle, which presents a typical clock reaction\cite{Lente-etal:NJC2007} effectively raising the pH level. Under suitable conditions, the clock recovers due to the exchange of acid and urea with an external reservoir, leading to
a \ch{pH} oscillator that differs from the wide class of conventional feedback oscillators\cite{Orban:ACR2015};
instead, it resembles a canard dynamics \cite{Benoit-etal:CM1981, Desroches-Jeffrey:PRSA2011}.
The insight gained into the oscillation mechanism can help to optimize experimental setups and to design chemical oscillators based on the same principles.

The presented stochastic analysis, in contrast to deterministic studies, shows that intrinsic noise
induces a significant statistical variation of the oscillation period, which increases upon downscaling the vesicle size. We note that although the mean period is remarkably robust for intermediate vesicle sizes, it slightly changes with the vesicle size. Therefore and because of the inevitable size disparity in vesicle suspensions \cite{Miele:Proc2016}, different oscillators possess slightly detuned eigenfrequencies, an important issue for understanding intervesicle communication and synchronization of rhythms\cite{Pikovsky-etal:2001book, Budroni-etal:JPCL2020, Budroni-etal-PCCP2021}, which would not be captured  by deterministic models.
Finally, our findings suggest that below a certain scale, which may be still relevant for applications, the periodicity of the rhythm is gradually destroyed. Namely, apart from the irregularity in the period length, there appear strong deviations in the oscillation amplitude masked by fluctuations growing with the decrease in vesicle size.
It is likely that similar trends take place for other \ch{pH} oscillators, which can be answered by specific tests along the lines presented here.

\begin{acknowledgement}
We thank Tam\'as B\'ans\'agi and Federico Rossi for clarifying details of the models in Refs.~\citenum{Bansagi:JPCB2014, Miele:LNBE2018}.
This research has been supported by Deutsche Forschungsgemeinschaft (DFG) through grant SFB~1114, project no.\ 235221301 (sub-project C03) and under Germany's Excellence Strategy -- MATH+ : The Berlin Mathematics Research Center (EXC-2046/1) -- project no.\ 390685689 (subproject AA1-1).
\end{acknowledgement}


\bibliography{references}

\clearpage

\onecolumn


\setcounter{page}{1}
\renewcommand*{\thepage}{S\arabic{page}}

\renewcommand*\suppinfoname{Supporting Information}
\begin{suppinfo}
		
\setcounter{figure}{0}
\renewcommand{\thefigure}{S\arabic{figure}}

{	
\setstretch{1.75}	

\subsubsection{Deterministic reaction rate equations}

\setcounter{equation}{0}
\renewcommand{\theequation}{S\arabic{equation}}

As suggested earlier \cite{Bansagi:JPCB2014}, the original eight-variable model can be simplified by neglecting the production of \ce{CO2}, leaving us with five species, see eqs.~(A2)--(A7) in Ref.~\citenum{Bansagi:JPCB2014}. From numerical tests we found that also the explicit dynamics of \ce{OH-} has no qualitative and no pronounced quantitative effect on the dynamics of the other species. Aiming at a minimal model capable of oscillations, we therefore neglect the presence of \ce{OH-}. Formally, this can be achieved by neglecting the exchange with the exterior for \ce{OH-} and making use of the quasi-steady state assumption\cite{Segel-Slemrod:SIAM-Rev1989}, assuming that the evolution of \ce{OH-} is fast compared to that of the catalytic step, \cref{4sm-R1} of the main text. Hence, setting ${\rm d}[\ce{OH-}]/{\rm d}t=0$ yields a slaved dynamics of the form  $[\ce{OH-}] \propto [\ce{H+}]^{-1}$.
As a result, we obtain the four-variable model discussed in the paper.
The temporal evolution of the urea--urease reaction scheme, see reactions \eqref{4sm} and \eqref{4sm-outp}, coupled to a reservoir according to \cref{24sm-exch2}
is governed by the following reaction rate equations for the mean concentrations:
\begin{subequations} \label{eq:rre}
	\begin{align}
	\frac{{\rm d}[\ce{S}]}{{\rm d}t} & = -k_\textrm{cat}([\ce{S}],[\ce{H+}])[\ce{S}]+k_\textrm{S}([\ce{S}_\textrm{ext}]-[\ce{S}])\,,  \\
	\frac{{\rm d}[\ce{H+}]}{{\rm d}t} & = k_{2r}[\ce{PH+}]-k_2[\ce{P}][\ce{H+}] + k_\textrm{H}([\ce{H+}_\textrm{ext}]-[\ce{H+}])\,, \\
	\frac{{\rm d}[\ce{P}]}{{\rm d}t} & = 2 k_\textrm{cat}([\ce{S}],[\ce{H+}])[\ce{S}] + k_{2r}[\ce{PH+}]-k_2[\ce{P}][\ce{H+}]-k [\ce{P}]\,,  \\
	\frac{{\rm d}[\ce{PH+}]}{{\rm d}t} & = k_2[\ce{P}][\ce{H+}] -  k_{2r}[\ce{PH+}] - k [\ce{PH+}]\,,	
	\end{align}
\end{subequations}
where the rate $k_\textrm{cat}$ of the catalytic step depends on
[\ch{S}] and [\ch{H+}] as given by \cref{eq:kcat,eq:kcat-MM} of the main text.
The mean molecule numbers shown in Figure~\ref{fig:evol-4sm-250nm} of the main text are obtained by rescaling with the molar volume, e.g., $X_{\ch{S}}=[S] V_\textrm{M}$ and accordingly for all other species.

\setcounter{figure}{0}
\renewcommand{\thefigure}{S\arabic{figure}}

\subsubsection{Breakdown of periodic rhythms upon downscaling}

As stated in the main text, below \SI{200}{\nano\meter} the rhythms are found to gradually lose its periodicity. In Figure~\ref{fig:evol-4sm-100nm}, we show the stochastic evolution for a \SI{100}{\nano\meter} vesicle, cf. Figure~\ref{fig:evol-4sm-250nm} of the main text.

}

\begin{figure*}
	\centering
	\includegraphics[width=0.7\textwidth]{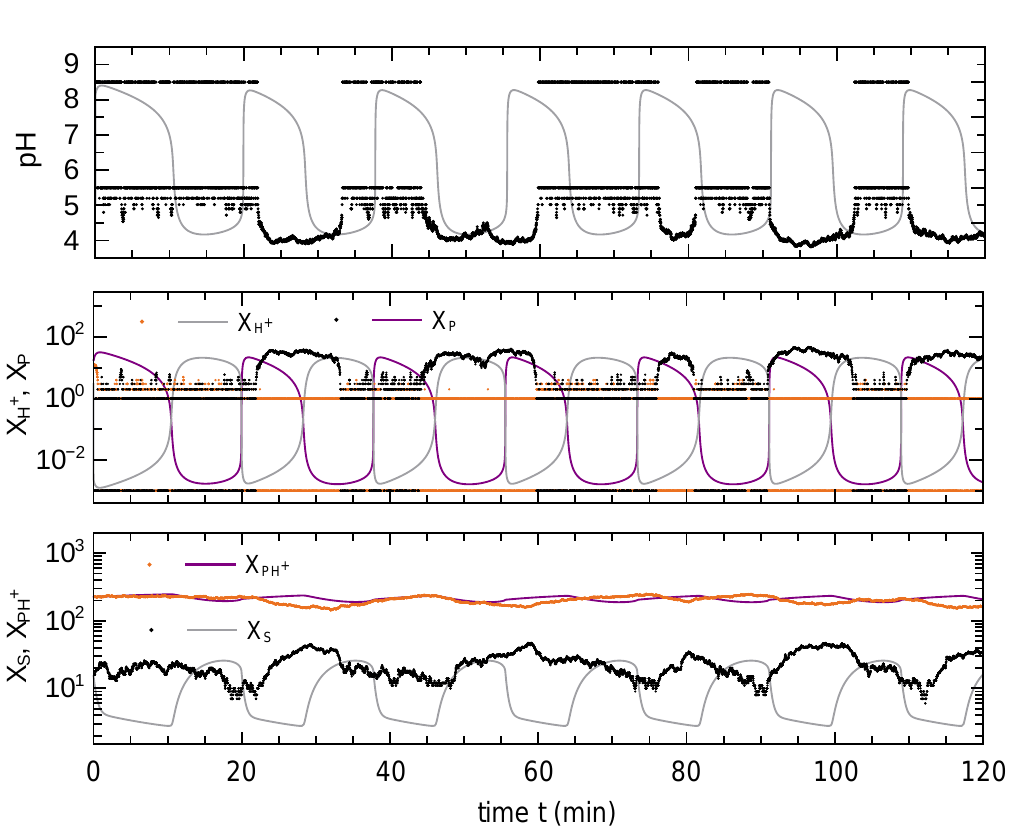}
	\caption{\textbf{Breakdown of periodic rhythms for a 100~nm vesicle} (reaction volume of $V=\SI{5.24e-19}{\liter}$).
		The \ch{pH} level (upper panel) and molecule numbers for the species \ch{H+} and \ch{P} (middle panel) and \ch{S} and \ch{PH+} (bottom panel) as functions of time for the urea--urease reaction scheme, see \cref{4sm,4sm-outp,24sm-exch2} of main text. Note the logarithmic scales.
		Solid lines are solutions to the deterministic reaction rate equations [Supporting Information, eqs.~\eqref{eq:rre}a--d].
		Dots show an exemplary solution to the stochastic reaction dynamics; the special value $X_{\ce{H+}} = 0$ is represented as $0.001$ ($\ch{pH}\approx 8.5$).
	}
	\label{fig:evol-4sm-100nm}
\end{figure*}

\end{suppinfo}

\end{document}